\documentclass[sigconf,screen]{acmart}

\setcopyright{acmlicensed}
\acmPrice{15.00}
\acmDOI{10.1145/3597926.3598128}
\acmYear{2023}
\copyrightyear{2023}
\acmSubmissionID{issta23main-p348-p}
\acmISBN{979-8-4007-0221-1/23/07}
\acmConference[ISSTA '23]{Proceedings of the 32nd ACM SIGSOFT International Symposium on Software Testing and Analysis}{July 17--21, 2023}{Seattle, WA, United States}
\acmBooktitle{Proceedings of the 32nd ACM SIGSOFT International Symposium on Software Testing and Analysis (ISSTA '23), July 17--21, 2023, Seattle, WA, United States}
\received{2023-02-16}
\received[accepted]{2023-05-03}

\ccsdesc[500]{Software and its engineering~Software testing and debugging}

\usepackage{float}
\usepackage{makecell}
\usepackage{enumitem}
\usepackage{natbib}
\usepackage{color}
\usepackage{graphicx}
\usepackage{listings}
\usepackage{siunitx}
\usepackage{microtype}
\usepackage{array}
\usepackage{tabularx}
\usepackage{hyperref}

\newcommand{\punt}[1]{}

\definecolor{mygreen}{rgb}{0,0.6,0}
\definecolor{mygray}{rgb}{0.5,0.5,0.5}
\definecolor{mymauve}{rgb}{0.58,0,0.82}
\makeatletter
\newcommand{\systemname}[1]{%
  \if@ACM@anonymous
    \textsc{Constrictor}%
  \else
    \textsc{SlipCover}%
  \fi
}
\newcommand{\ifanonymous}[1]{%
  \if@ACM@anonymous #1\fi
}
\newcommand{\ifnotanonymous}[1]{%
  \if@ACM@anonymous\else #1\fi
}
\makeatother

\begin{document}

  \title{\systemname{}: Near Zero-Overhead Code Coverage for Python}

  \author{Juan Altmayer Pizzorno}
  \email{jpizzorno@cs.umass.edu}
  \affiliation{%
    \institution{College of Information and Computer Sciences \\
                 University of Massachusetts Amherst}
    \country{United States}
  }
  \author{Emery D. Berger}
  \email{emery@cs.umass.edu}
  \affiliation{%
    \institution{College of Information and Computer Sciences \\
                 University of Massachusetts Amherst}
    \country{United States}
  }

  \begin{abstract}
    Coverage analysis is widely used but can suffer from high overhead.
This overhead is especially acute in the context of Python, which is
already notoriously slow (%
a recent study
observes a roughly 30$\times$
slowdown vs.  native code).  We find
that the state-of-the-art coverage tool for
Python, \texttt{coverage.py}, introduces a median overhead of 180\%
with the standard Python interpreter.
Slowdowns are even more extreme when using PyPy, a
JIT-compiled Python implementation, with \texttt{coverage.py} imposing a
median overhead of 1,300\%.
This performance degradation reduces the utility of coverage analysis in most
use cases, including testing and fuzzing, and precludes its use in
deployment.

This paper presents \systemname{}\ifanonymous{\footnote{Name altered for double-blind review.}}, a novel, near-zero overhead coverage
analyzer for Python. \systemname{} works without modifications to either the Python interpreter or PyPy.
It first processes a program's AST to accurately identify all branches and lines.
\systemname{} then dynamically rewrites Python bytecodes to add lightweight
instrumentation to each identified branch and line.  At run
time, \systemname{} periodically \emph{de-instruments} already-covered
lines and branches. The result is extremely low overheads---a median
of just 5\%---making \systemname{} suitable for use in deployment. We
show its efficiency can translate to significant increases in
the speed of coverage-based clients. As a proof of concept, we
integrate \systemname{} into TPBT, a targeted
property-based testing system, and observe a 22$\times$ speedup.

  \end{abstract}

  \keywords{Testing, Code Coverage, Dynamic Code Instrumentation, Python}

  \maketitle
    
  \begin{figure*}[t]
    \includegraphics[width=\textwidth]{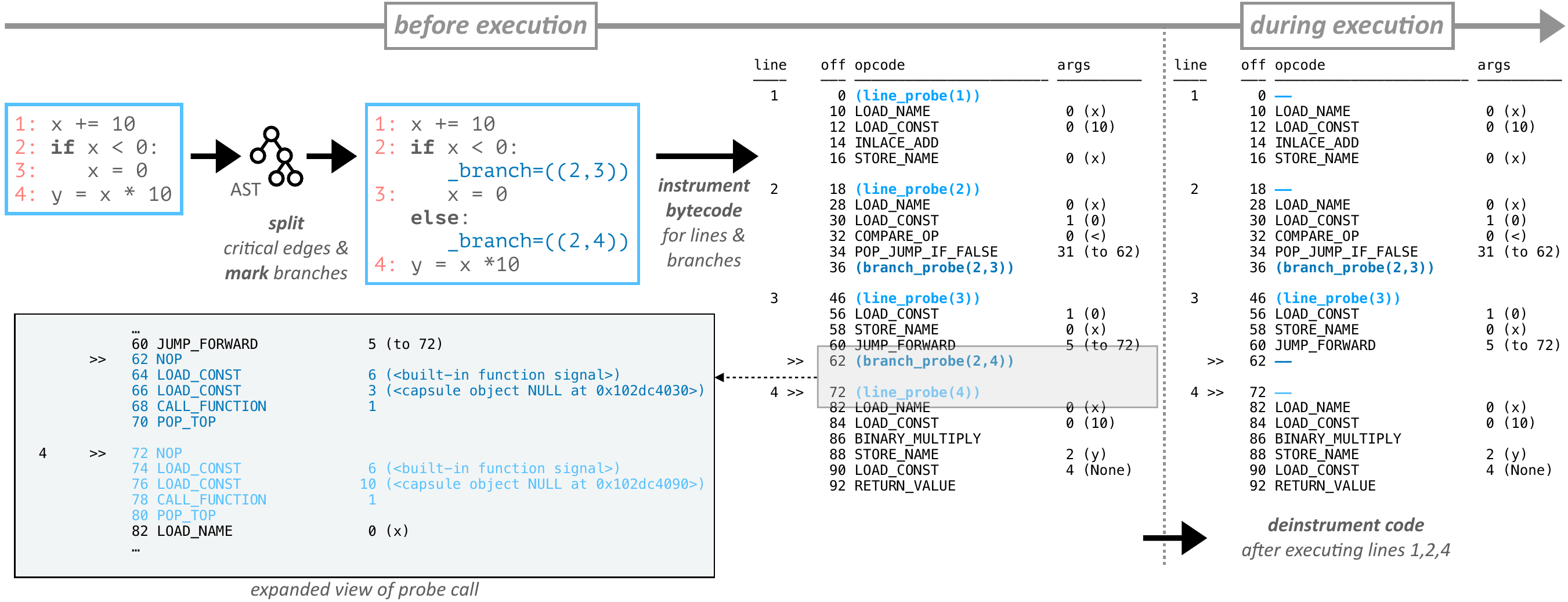}
    \vspace{0.5em}
    \caption{
        \textbf{Overview:} \systemname{} drastically reduces overhead compared to the state-of-the-art tracing-based coverage analysis tool for Python (\texttt{coverage.py}) via a combination of three approaches: (1) AST transformation (Section~\ref{sec:overview-ast-transformation}), (2) lightweight bytecode instrumentation (Section~\ref{sec:overview-bytecode-instrumentation}), and (3) dynamic bytecode de-instrumentation (Section~\ref{sec:overview-de-instrumentation}). \textnormal{AST transformations let \systemname{} accurately identify all branches in the program source. Inserted bytecode instrumentation then tracks lines and branches at runtime. In code regions where lines or branches have been covered, \systemname{} gradually rewrites their bytecode to eliminate instrumentation, letting the Python code subsequently run at full speed.
Python bytecode varies with version. This paper uses Python~3.10 bytecode throughout.}
        \label{fig:overview}
    }
\end{figure*}

  \section{Introduction}
  \label{sec:introduction}


Code coverage analysis is widely used for a variety of tasks including
testing, program analysis such as slicing, and statistical fault localization.
Branch and line coverage information can reveal areas of
code that are not exercised by a test suite. Automated testing
approaches, like mutation testing and fuzzing, use coverage
information to evaluate mutants or drive the exploration of new code
paths.


Unfortunately, existing code coverage analyzers often substantially degrade
performance~\cite{DBLP:journals/stvr/ChilakamarriE06}.
The source of this slowdown is primarily due to overhead that the dynamic analysis
imposes to obtain the coverage information at execution
time~\cite{DBLP:journals/stvr/ChilakamarriE06, DBLP:journals/jss/TikirH05}.
Code coverage analyzers typically either instrument each line of code,
inserting \emph{probes} that record when these are reached, or, in the
case of Python, build on that language's built-in tracing support. 
The slowdown imposed by code coverage analyzers has a significant
impact not only on the time to run tests or the number of tests that
can be run in a given time
budget. At Google, ``coverage instrumentation is relatively expensive, and
performance failures are the most common cause of failed coverage
computations''~\cite{DBLP:conf/sigsoft/IvankovicPJF19}.


In the context of Python, which is already notoriously
slow~\cite{DBLP:conf/usenix/LionCSY22}, coverage analysis leads
to dramatic slowdowns. As Section~\ref{sec:evaluation} shows, the popular and widely used
state-of-the-art code coverage analysis tool for
Python, \texttt{coverage.py} (used at Google, with 7.4 million
downloads per week)~\cite{coveragepy}, increases execution times by
between 30\% and 260\% (median: 180\%) vs. CPython,
the standard Python interpreter~\cite{cpython}. This overhead is
even more extreme when running PyPy, a JIT compiled implementation of
Python~\cite{pypy}, slowing down programs by 140\%--32,400\% (median: 1,300\%).


\punt{
Previous work aimed at reducing the cost of coverage has proposed
removing probes once they execute. Since lines and branches only need to
be marked as covered once, removing now-redundant probes allows
subsequent execution of code to run at full speed.  Previous work
explored this idea for Java~\cite{residual,
DBLP:journals/stvr/ChilakamarriE06, Misurda1553558} and for native
code~\cite{DBLP:journals/jss/TikirH05, DBLP:conf/sp/NagyH19}.
Unfortunately, these previous approaches are unsuitable for Python, as
they either operate at the wrong level of abstraction or would require
changes to each language implementation; we also find that some
previous work would be unsound in modern language implementations
(see Section~\ref{sec:related-work}).
}


This paper presents \systemname{}, a near-zero overhead coverage
analysis approach for Python (CPython median: 5\%).
\systemname{} first performs an AST transformation to
accurately identify branches in the Python source code.  It then
inserts low-overhead instrumentation
that tracks branch and line coverage at runtime.
Critical to its efficiency, it gradually
eliminates probes as lines or branches are covered, eventually letting
code run at full speed. \systemname{} manages the cost of
de-instrumentation with its expected performance benefits, triggering
de-instrumentation only once it crosses a cost-benefit threshold.

We show that \systemname{} accurately collects coverage information
while effectively eliminating the overhead of coverage analysis.
\systemname{}'s efficiency not only can drastically reduce time spent testing
and expand the number of fuzzing, mutation tests, or unit tests in a
given time budget.  It also enables new use cases for coverage
analysis: for example, because of its low overhead, \systemname{}
could be used in deployment to identify code that is never used by any
client, and is thus suitable for
removal~\cite{DBLP:journals/jss/TikirH05}. In a proof of concept, we
show the value of speeding coverage analysis in the context of a
coverage-guided property-based testing system~\cite{tpbt}, where it
yields a $22\times$ speedup.

This paper makes the following contributions:
\begin{itemize}[topsep=0pt,leftmargin=*]
    \item It presents \systemname{}, a new line and branch coverage analyzer for Python based on AST transformation,
        bytecode instrumentation and de-instrumentation.
    \item It conducts a detailed empirical analysis of \systemname{}, showing that it significantly advances the state of the art, almost completely eliminating the overhead of coverage analysis.
\end{itemize}

  \section{\systemname{} Overview}
  \label{sec:overview}

Figure~\ref{fig:overview} presents a graphical overview of \systemname{}.
\systemname{} operates in two phases: a static phase and a dynamic phase.
The static phase, shown to the left of the dotted line, performs AST transformations (Section~\ref{sec:overview-ast-transformation})
and inserts probes as the Python interpreter loads code (Section~\ref{sec:overview-bytecode-instrumentation}).
The dynamic phase tracks coverage and dynamically eliminates probes when they are no longer needed (Section~\ref{sec:overview-de-instrumentation}).

The figure shows on the left a sample portion of Python code, followed by the results of
\systemname{}'s AST transformation and the resulting instrumented bytecode.
On the right, it shows the same bytecode after \systemname{} has partially de-instrumented it.

\subsection{AST Transformation}
\label{sec:overview-ast-transformation}

\begin{figure}[t]
    \includegraphics[width=.3\columnwidth]{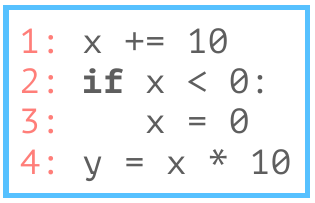}
    \caption{
        \textbf{Example:} An \texttt{if} statement is used without a matching \texttt{else}.
        \label{fig:if-without-else}
    }
\end{figure}

\begin{figure}[t]
    \includegraphics[width=.8\columnwidth]{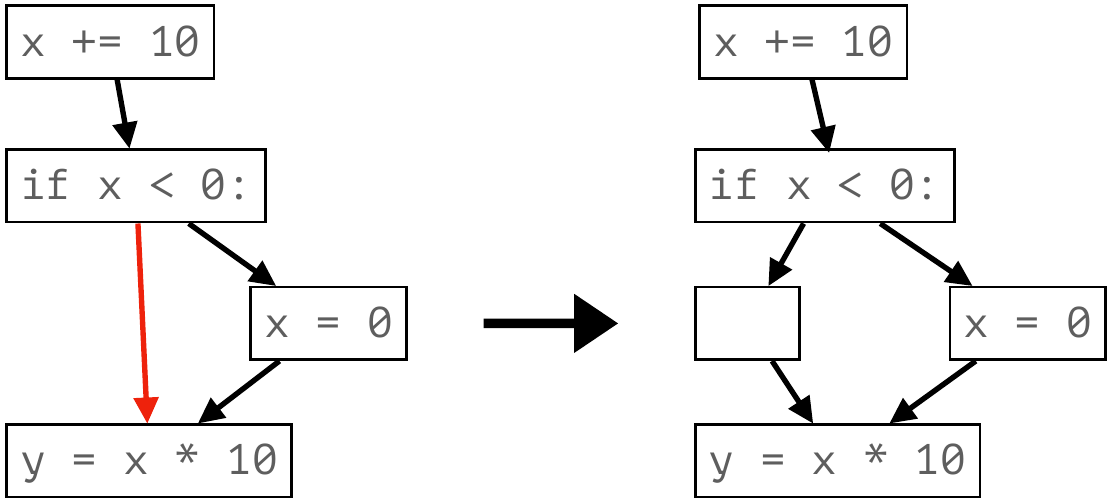}
    \caption{
        \textbf{Critical Edge:} \systemname{} transforms the code example's AST,
        spliting a critical edge (shown in red) to facilitate efficient instrumentation.
        \label{fig:critical-edge}
    }
\end{figure}

\systemname{}'s static phase begins with AST transformations to split \emph{critical edges}
in the control flow graph and to mark branches.  Critical edges result
from statements that are optional or empty.
As an example, Figure~\ref{fig:if-without-else} shows a code excerpt containing an
\texttt{if} statement without a matching \texttt{else}, and Figure~\ref{fig:critical-edge}
the resulting control flow graph.

Branch coverage analysis requires that the system record whether
any branch was taken, even if one of them is implicit.
Splitting these edges lets \systemname{} perform branch coverage
analysis accurately and efficiently using bytecode
instrumentation. \systemname{} parses the code and transforms the
resulting abstract syntax tree to add explicit branches for these
cases. To simplify this transformation, \systemname{} leverages an
idiosyncrasy of Python, which has explicit \texttt{else} statements
not only for \texttt{if} but also for most control flow constructs
(e.g.,
\texttt{while} and \texttt{for}).


\systemname{} then marks the branches, inserting placeholder assignments
which the instrumentation step replaces with probe calls.
Figure~\ref{fig:branch-if} shows the resulting transformation.

\subsection{Bytecode instrumentation}
\label{sec:overview-bytecode-instrumentation}

The final step in the static phase performs bytecode instrumentation.
\systemname{} uses source line information and any branch placeholders
to guide the insertion of probes in the code.
The probes perform three roles: (1) they mark the relevant line or branch ``covered'',
(2) they track the number of times they execute, and (3) they trigger de-instrumentation
once that counter crosses a threshold.

\systemname{}'s probes consist of two parts: a C++ extension module, and bytecode sequences
(shown in expanded view in Figure~\ref{fig:overview}) that call into
that module.  This two-pronged approach has two benefits. First,
placing most of the probe functionality in a native extension module
helps keep the probes lightweight, avoiding Python interpreter
overhead and leveraging Python's efficient integration with native
code. Second, instrumenting the bytecode to call into it
lets \systemname{} avoid having to modify the interpreter.  As
Section~\ref{sec:implementation-de-instrumenting-probes} describes,
this approach also helps keep de-instrumentation efficient.

\subsection{Program execution and dynamic de-instrumentation}
\label{sec:overview-de-instrumentation}

Once \systemname{} finishes instrumenting the program, it initiates
program execution, marking the beginning of the dynamic phase. If the
program imports (loads) additional code during
execution, \systemname{} suspends the dynamic phase, performs its
static phase steps on the new code, and then resumes execution.

Now, as the program being executed reaches each line and branch for
the first time, it first executes the inserted probe. The first time a
probe is executed, it records the line or branch as being covered, which
implicitly marks the probe as eligible for eventual elimination (rather than
immediately eliminating it, for reasons discussed in the next paragraph). Subsequent
executions of the probe only increment a counter and check to see if
it crosses a predefined threshold count. Upon crossing that threshold,
the probe triggers elimination of all marked probes.

Delaying de-instrumentation makes \systemname{} more efficient.
Bytecode modifications are relatively expensive, and batching
de-instrumentation amortizes this overhead. This approach also
lets \systemname{} limit de-instrumentation costs for rarely executed
code, whose probe cost yields limited overhead. If probes execute
often, more are eliminated, but if they only execute once, or even a few
times, no time is wasted eliminating them.

  \section{\systemname{} Implementation}
  This section describes key portions of \systemname{}'s implementation in detail,
including its AST transformation for branch coverage (Section~\ref{sec:implementation-branch-coverage}), probe insertion (Section~\ref{sec:implementation-instrumenting-bytecode}), probe elimination (Section~\ref{sec:implementation-de-instrumenting-probes}), and code loading interception (Section~\ref{sec:implementation-intercepting-code-loading}).

\subsection{\textbf{Instrumenting for Branch Coverage}}
\label{sec:implementation-branch-coverage}
As Section~\ref{sec:overview} describes, the first step in \systemname{}'s static phase
performs AST transformations to identify the branches in the code.
As an alternative, we considered performing control-flow analysis on the bytecode,
identifying branches through jump instructions and their source and
destination lines.
One advantage of this approach would have been that \systemname{} would
be able to operate without source code.
While Python code is generally distributed in source form, the
bytecode can be executed without accompanying source code.

Unfortunately, this approach is unworkable. Python's bytecode
compiler performs transformations that make the correspondence between
jump instructions and their branches in the source code difficult if
not impossible to identify. This challenge was first identified for
coverage tools that rely exclusively on Java bytecode by \citet{DBLP:conf/issre/LiMOD13},
who found that the exclusive reliance on bytecode leads to inaccurate branch coverage analysis.

\begin{figure}[!t]
    \includegraphics[width=.8\columnwidth]{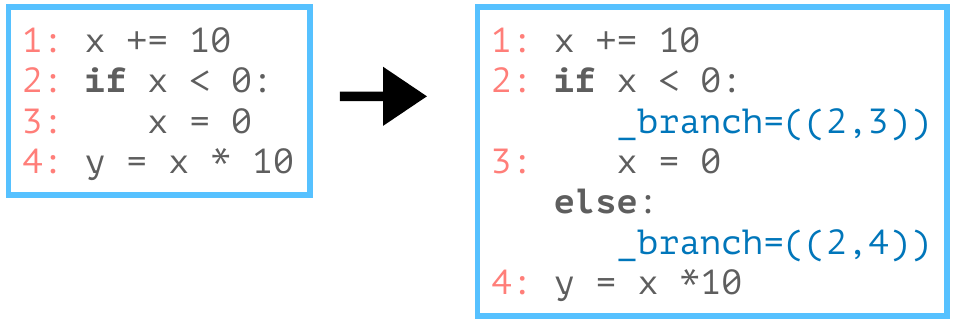}
    \caption{\textbf{Demarcating branches:} \systemname{} inserts assignment statements to demarcate branches.
        \label{fig:branch-if}
    }
\end{figure}

Instead, \systemname{} employs a hybrid approach to collecting branch coverage.
It uses Python's compiler and \texttt{ast} library to parse Python source into an
AST.
It then transforms the AST, splitting critical edges (see Figure~\ref{fig:critical-edge})
and demarcating all branches by injecting assignment statements into the code
that encode the origin and destination lines of each branch
(e.g., \texttt{\_branch = (origin,~dest)}).
Unlike branches, assignment statements are straightforward to recognize in
generated bytecode.
Figure~\ref{fig:branch-if} shows this transformation for an \texttt{if} statement.

\begin{figure}[!t]
    \includegraphics[width=\columnwidth]{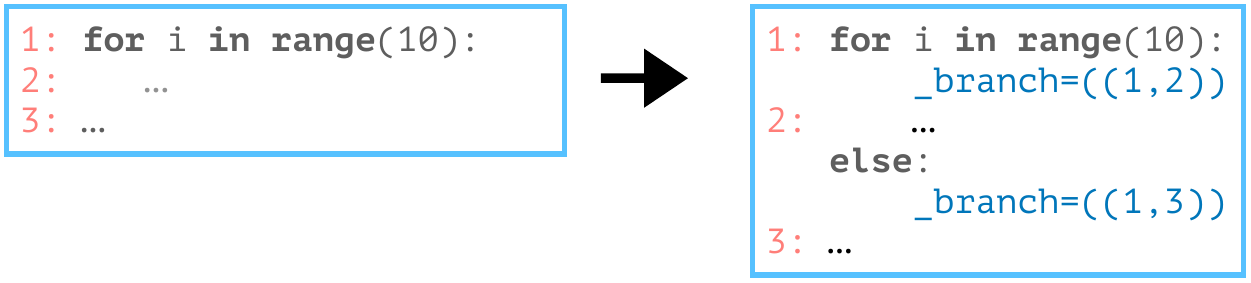}
    \caption{\textbf{Demarcating loop branches:}
        Unusually, Python allows the use of \texttt{else} with \texttt{for} and \texttt{while} statements,
        simplifying inserting code to demarcate the block skipping branch
        ($1\rightarrow3$ in the example).
        \label{fig:instrumenting-loops}
    }
\end{figure}

\systemname{}'s AST transformation pass splits critical edges, making all branches explicit.
It leverages an idiosyncrasy of Python that simplifies this AST
transformation.  Unlike other widely-used programming languages,
Python's \texttt{else} statement not only combines with \texttt{if},
but also with \texttt{for} and \texttt{while} statements. In these
cases, the \texttt{else} branch is executed when the
conditional evaluates to \texttt{False}. To make all branches
explicit, \systemname{} simply inserts an \texttt{else} statement,
when missing, after all \texttt{if}s and \texttt{for} and \texttt{while} loops.
Figure~\ref{fig:instrumenting-loops} shows an example of this
transformation.

\begin{figure}[!t]
    \includegraphics[width=\columnwidth]{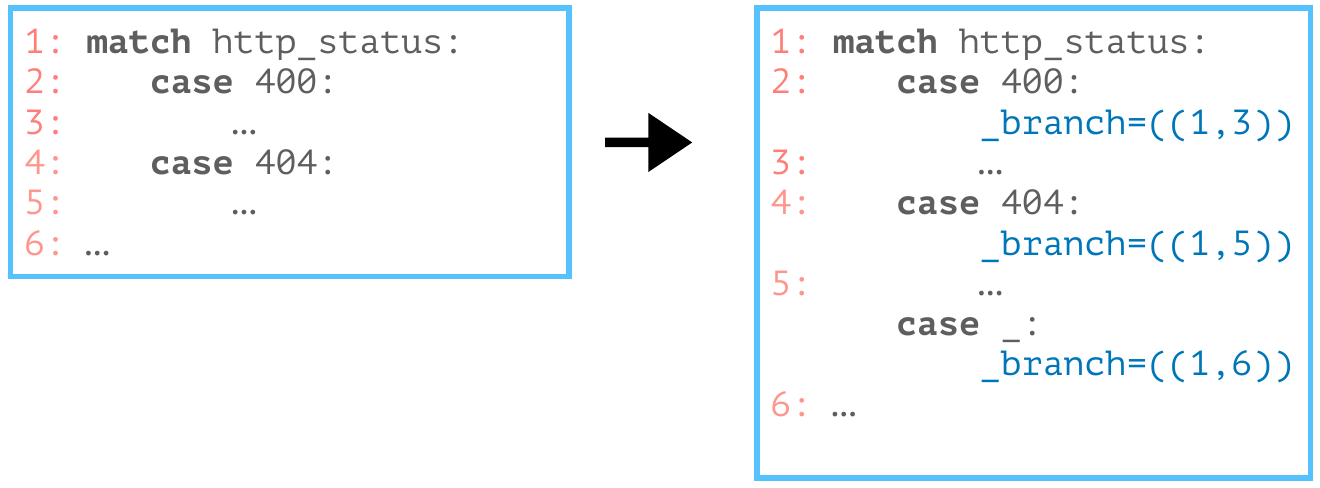}
    \caption{\textbf{Demarcating \texttt{match}:} \systemname{} inserts a ``\texttt{match~\_}''
        statement, if missing, to demarcate that branch.
        \label{fig:branch-match}
    }
\end{figure}

Python~3.10 introduced \texttt{match} statements~\cite{python-match-statement},
which are structurally similar to the well-known C \texttt{switch} statement.
While \texttt{match} cannot be combined with \texttt{else}, its wildcard
\texttt{case~\_:} statement can be used similarly.
Figure~\ref{fig:branch-match} shows an example.

Having performed these modifications, \systemname{} then compiles the AST into bytecode.
The next step in the static phase replaces the demarcations with branch probes.

\subsection{\textbf{Inserting Probes}}
\label{sec:implementation-instrumenting-bytecode}

\begin{figure}[!t]
    \includegraphics[width=\columnwidth]{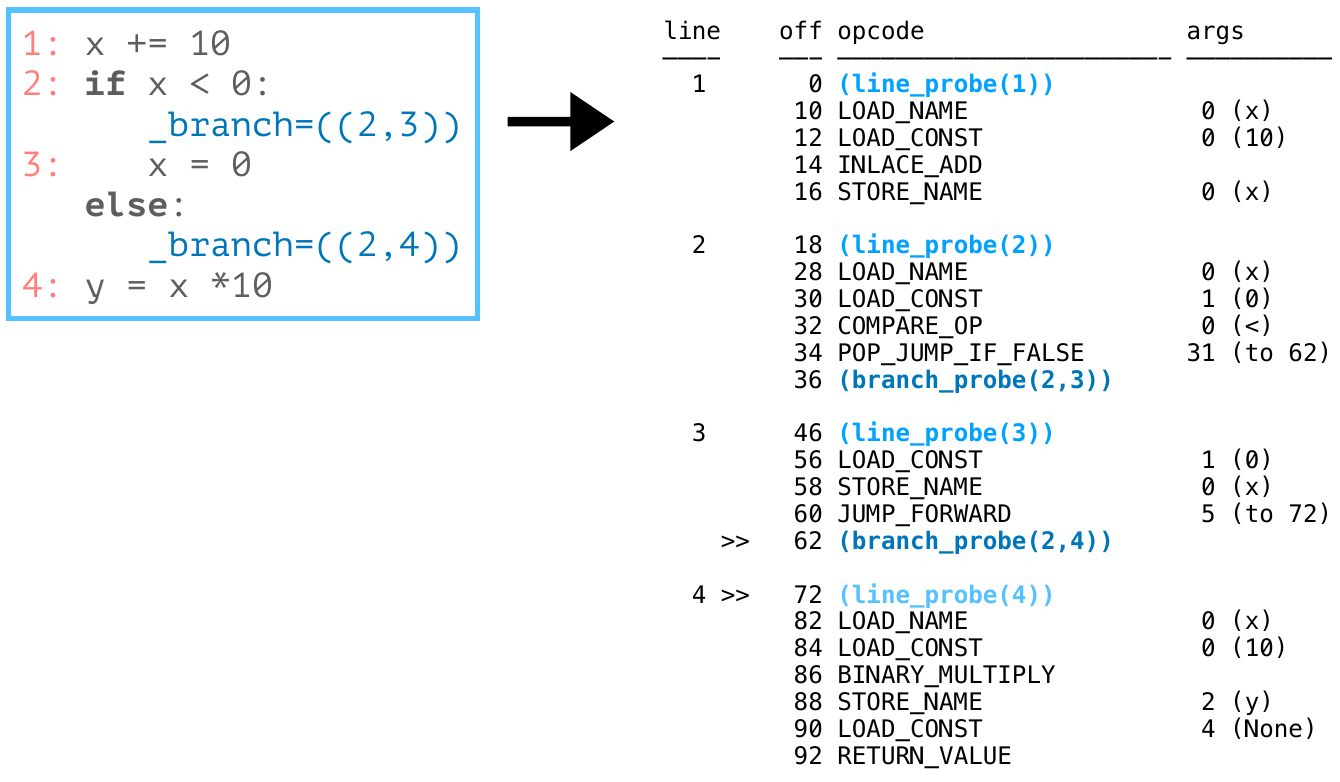}
    \caption{
        \textbf{Inserting Probes:} \systemname{} inserts line and branch probes
        guided by Python metadata and its own branch demarcations.
        \textnormal{For clarity, \texttt{line\_probe($l$)} and \texttt{branch\_probe(($l_1,l_2$))} stand
        for the actual probe bytecode sequences (see Figure~\ref{fig:probe-bytecode}).}
        \label{fig:probe-insertion}
    }
\end{figure}

The second and last step in \systemname{}'s static phase insert probes into
the bytecode.
This step also reads from a bytecode object, which is either the one compiled from the
AST in the first static step, or the unmodified bytecode loaded from Python, if
no branch coverage is being collected.
Python bytecode objects include metadata indicating the source code file and line
corresponding to each portion of code.
\systemname{} uses this information to determine where to insert
its line probes.
\systemname{} scans the code for the special \texttt{\_branch}
assignments introduced in the first step (Section~\ref{sec:implementation-branch-coverage}).
As Figure~\ref{fig:probe-insertion} shows, \systemname{} inserts probes ahead of the code for
each line and branch.

\begin{figure}[!t]
    \includegraphics[width=\columnwidth]{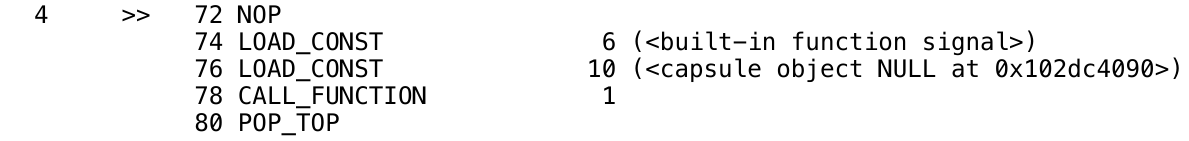}
    \caption{
        \textbf{Probe bytecode:} \texttt{line\_probe(4)} sequence from Figure~\ref{fig:probe-insertion}.
        \textnormal{\systemname{} replaces the \texttt{NOP} opcode with a \texttt{JUMP\_FORWARD}
        to skip over the entire sequence when eliminating the probe.}
        \label{fig:probe-bytecode}
    }
\end{figure}


Figure~\ref{fig:probe-bytecode} presents an actual probe bytecode
sequence. These start with a \texttt{NOP} opcode
that \systemname{} later may replace with a \texttt{JUMP\_FORWARD} to
skip over (eliminate) the probe sequence (see
Section~\ref{sec:implementation-de-instrumenting-probes}).  The rest
of the sequence invokes the native portion of the probe, which is
wrapped in a Python ``capsule'' object.
\systemname{} instantiates a new capsule object for each line and branch probe
as it inserts them.
Since their addresses are known at insertion time, \systemname{} stores all addresses
as code constants, avoiding the need to perform costly lookups during execution.


The native portion of the probe implements the rest of the probe logic.
It records the newly covered line or branch when it is first executed, and triggers
de-instrumentation once it has executed a certain number of times
(Section~\ref{sec:implementation-de-instrumenting-probes}).

\begin{figure}[!t]
    \includegraphics[width=\columnwidth]{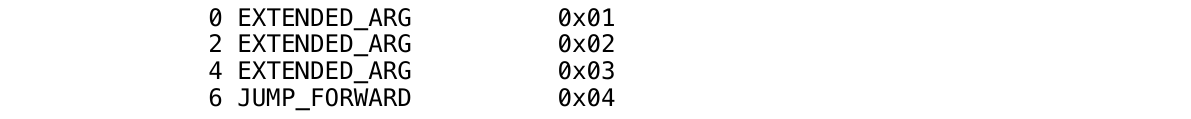}
    \caption{
        \textbf{Extended arguments:} Python uses a special
        \texttt{EXTENDED\_ARG} opcode to encode multi-byte arguments
        \textnormal{(a jump forward to \texttt{0x01020304} in this case)}.
        \label{fig:extended-arg}
    }
\end{figure}

Modifying Python bytecode is one of the more intricate parts of \systemname{}.
Since Python bytecode objects are read-only, \systemname{} must create new
objects to insert probes.
Python opcodes also have variable length. The base form is just one byte
for the opcode and one for the argument, but it may also be preceded by up to
three \texttt{EXTENDED\_ARG} opcodes if the argument is longer than one byte.
Figure~\ref{fig:extended-arg} shows an example.


\systemname{} inserts probes by shifting the original bytecode down
at each insertion point.
As it does so, it updates existing jump destinations and
various metadata, including the line table and the exception handler table.
Unfortunately, inserting bytecode may also increase the offset (or difference
in offset, for relative jumps) between jump opcodes and their destinations.
If the new destination then does not fit in the space occupied by the
original jump, \systemname{} may need to insert additional \texttt{EXTENDED\_ARG} opcodes.
Inserting \texttt{EXTENDED\_ARG}s may require updating the jumps again.
\systemname{} deals with this issue via the same approach as the CPython compiler:
it repeatedly updates jumps as long as any length adjustments are necessary.

It might seem possible to adapt a design similar to that employed
by \citet{DBLP:journals/jss/TikirH05} in order to reduce
instrumentation overhead. Adapting their approach would involve replacing
the bytecode at each insertion point by a jump to a \emph{trampoline}
section of bytecode. This trampoline would be allocated at the end of
the original code. It would first call to the native probe, then
execute the bytecode that was replaced, and finally jump back to
the original code right after the insertion point. By avoiding
bytecode shifts, this approach theoretically offers the promise of
reducing probe insertion overhead. Moreover, it would also would have made it possible to
efficiently fully remove probes by restoring the original bytecode,
overwriting the jump to the trampoline.

Unfortunately, that approach is not viable in the context of Python. A
jump to a trampoline would not necessarily fit in the space between
two insertion points. For example, a line of Python code,
when compiled, might yield only two bytes: an opcode and its
argument. Code for the next line would then start at the third byte.
If a jump to the trampoline required four or more bytes (a likely scenario),
there would be no room to insert it without corrupting the following line's bytecode.

\subsection{\textbf{Eliminating Probes}}
\label{sec:implementation-de-instrumenting-probes}


\systemname{} structures its probes so that it can
eliminate probe overhead during de-instrumentation in a streamlined
manner, as this section describes.

\systemname{} records coverage in two sets: a ``known'' set and a ``newly covered'' set.
The first time a probe executes, it adds coverage information (e.g., this line is now covered) to
the newly covered set, which also indicates that the probe is ready for
elimination.

The first step in \systemname{}'s de-instrumentation is to set a flag
in the probe that indicates it should skip recording coverage
information in subsequent executions. This flag avoids much of the
overhead of instrumentation (adding to covered sets). Subsequent
executions only perform that check and then increment an execution
counter. Once the execution counter crosses a pre-defined threshold, it
performs the final de-instrumentation step, eliminating any previously
reached probes.

\systemname{} then atomically swaps the ``newly covered'' for a new, empty set,
where any additional coverage is recorded while it processes the items
previously in ``newly covered''.
It adds these items to the ``known'' set and then uses maps it created during
instrumentation to efficiently locate the bytecode objects containing their
respective probes.


As Section~\ref{sec:implementation-instrumenting-bytecode} describes, modifying bytecode
can be a complex operation.
To simplify and save time during probe elimination, \systemname{} does not delete
the probe sequences from bytecode. Instead, it inserts a jump over them, effectively
removing the probe without the complexity entailed by actually deleting the code (including
updating line numbers and jumps).

\systemname{} leverages another idiosyncratic behavior of Python to simplify this operation.
\systemname{}'s probes start with a \texttt{NOP} opcode.
All opcodes in Python are encoded with an argument, including \texttt{NOP}, whose argument Python ignores.
\systemname{} exploits this encoding to pre-encode the length of the probe sequence
into \texttt{NOP}'s argument. To eliminate a
probe, \systemname{} then simply replaces that
\texttt{NOP} with a \texttt{JUMP\_FORWARD} opcode.

Since Python bytecode objects are read-only, \systemname{} creates new
instances for any objects with newly eliminated probes.  Bytecode
objects are also tree-structured, as any nested functions, lambda
functions, or comprehensions are included as constants in the
enclosing bytecode object.

Consequently, to eliminate probes, \systemname{} performs a
depth-first search of the applicable bytecode objects, replacing these
bytecode ``constants'' as it moves from a child node to a parent node.
Python stores additional references to bytecode objects in
dictionaries that it uses to implement its various variable contexts.
Once all covered probes have been eliminated, \systemname{} searches
for any such references, replacing them with the new instances.
\systemname{}'s search is comprehensive, including all modules it has instrumented,
as well as every Python frame's global and local contexts.
Such updates are most efficiently performed in batches to reduce the frequency of searches.
This fact in part motivates \systemname{}'s threshold-based de-instrumentation.

\systemname{} thus gradually de-instruments the program as it executes, allowing it
to continue to run at close to full speed, while leaving as-yet unreached probes in place
to record coverage should an execution path reach them.

\punt{
While dynamic probe removal has the potential to lower overhead, it is not always
worthwhile, as it depends on the cost of removal as well as on the extent to which that portion of
the code executes again.
If the probe removal has nontrivial cost, it should ideally be managed, such as performing
it only on code that is likely to re-execute.
}

\subsection{\textbf{Intercepting Code Loading}}
\label{sec:implementation-intercepting-code-loading}

In order to instrument bytecode, \systemname{} must intercept code as the
Python interpreter loads it, which can happen at any time during execution.
It performs this task by inserting its own specialization of \texttt{importlib}'s
\texttt{MetaPathFinder} at the head of Python's import path.
This approach lets \systemname{} observe and interpose on all subsequent module load requests.
To reduce overhead, \systemname{} does not instrument every module,
but only that are deemed relevant (either by a default set, or those specifically specified by the user).

One complication is that the widely-used testing framework \texttt{pytest}
 also inserts its own \texttt{Loader}, which it uses to modify
unit tests in order to better support \texttt{assert} statements.
That loader would normally bypass \systemname{}'s, causing unit
tests to be excluded from the coverage measurement.
To avoid that, \systemname{} preloads \texttt{pytest}'s assertion rewriter and
``monkey patches'' it, interposing itself so it can instrument the unit tests
while still allowing \texttt{pytest} to later make its own changes.

  \section{Evaluation}
  \label{sec:evaluation}
Our evaluation answers the following research questions:
\begin{itemize}[topsep=0pt,leftmargin=*]
\item {\textbf{RQ1:}} How does \systemname{}'s performance compare to the state-of-the-art (\texttt{coverage.py}) (Section~\ref{sec:evaluation-rq1})?
\item {\textbf{RQ2:}} How much does \systemname{}'s de-instrumentation contribute to its high performance (Section~\ref{sec:evaluation-rq2})?
\item {\textbf{RQ3:}} How does \systemname{}'s approach behave across recent Python versions  (Section~\ref{sec:evaluation-rq3})?
\item {\textbf{RQ4:}} What is the overhead of approaches based on Python's tracing mechanism, as used by \texttt{coverage.py} (Section~\ref{sec:evaluation-rq4})?
\item {\textbf{RQ5:}} What is the impact of de-instrumentation when running in the PyPy JIT compiler (Section~\ref{sec:evaluation-rq5})?
\item {\textbf{RQ6:}} How does \systemname{} affect the performance of automated testing frameworks (Section~\ref{sec:evaluation-rq6})?
\end{itemize}

\subsection{Experimental Setup}
We perform all experiments on a 10-core 3.70GHz Intel Core~i9-10900X
system equipped with 64GB of RAM and an NVIDIA GeForce RTX~2080 SUPER
GPU, running Linux~5.16.19-76051619-generic.  We
compile \systemname{}'s C++ code as well as the various CPython
interpreter versions with GCC~9.4.0, utilizing Python's default flags,
which include
\texttt{-O3} optimization.
We utilize the stock PyPy~3.9-v7.3.11 distribution from \texttt{pypy.org}.
Except where noted otherwise, we use CPython~3.10.5 and \texttt{coverage.py}~6.4.4.
We perform all measurements on an otherwise quiescent system.
All measurements are taken five times, and we report the median.

We select as benchmarks a mix of test suites for
real-world applications, as well as compute-intensive Python
applications. The benchmarks include the test suites for the
scikit-learn~\cite{scikit-learn} and Flask~\cite{flask} packages, both
``real world'' users of coverage analysis, selected for relevance as
well as for their range of running times. Scikit-learn's suite runs
for about ten minutes (using CPython) and is representative of a long-running suite.
Flask's less comprehensive suite, whose runtime we extend by utilizing the
\texttt{pytest-repeat} package to repeat the tests five times, runs for less than ten seconds.
We also include several pure Python
applications: \texttt{fannkuch}, \texttt{mdp}, \texttt{pprint}, \texttt{raytrace}, \texttt{scimark}
and \texttt{spectral\_norm}, the longest-running benchmarks in the
standard Python Benchmark Suite used to evaluate Python
performance~\cite{pyperformance}, and Peter Norvig's Sudoku
solver~\cite{sudoku}, a compute-intensive pure Python application.

\subsection{[RQ1] How does \systemname{}'s performance compare to \texttt{coverage.py}?}
\label{sec:evaluation-rq1}

We compare \systemname{}'s and \texttt{coverage.py}'s performance by using these to analyze coverage
on our benchmark suite. We normalize results against their base running time, i.e., running
without any coverage analysis.

\begin{figure}[!t]
    \includegraphics[width=\columnwidth]{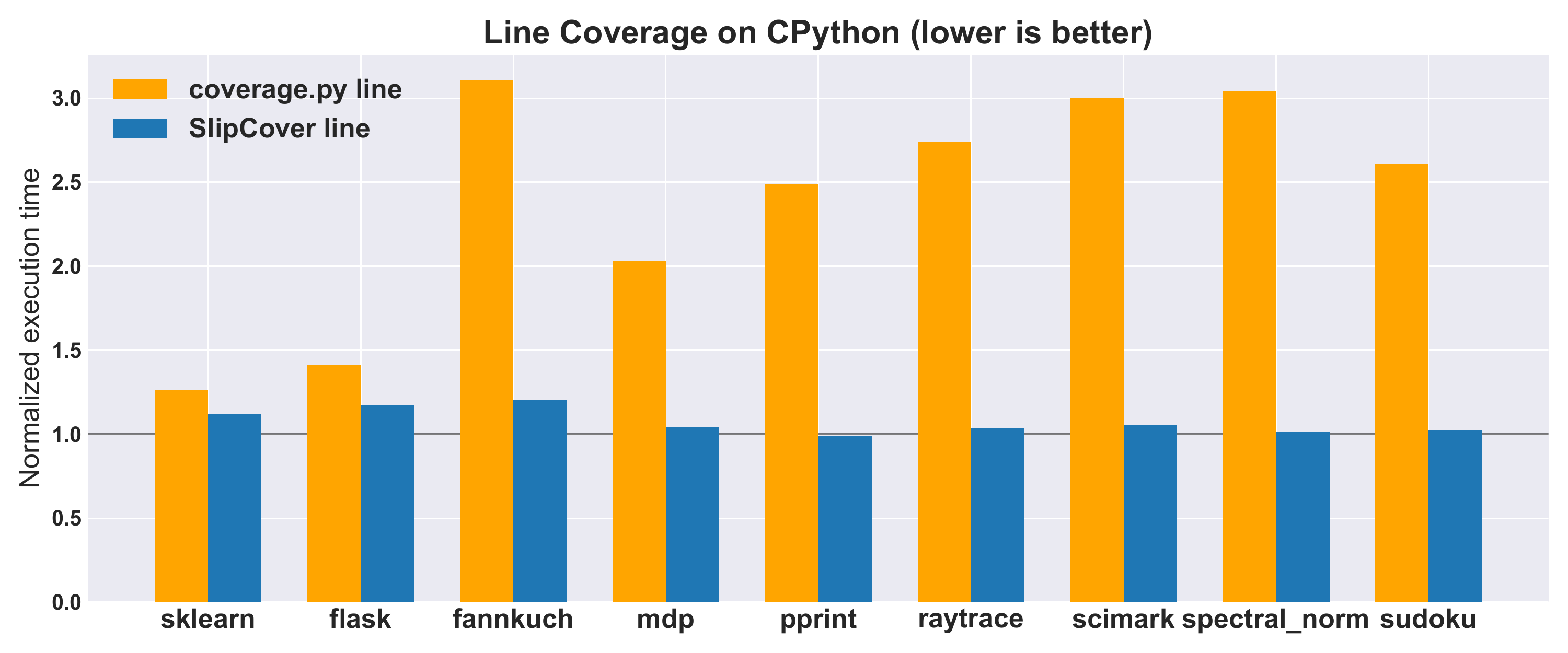}
    \includegraphics[width=\columnwidth]{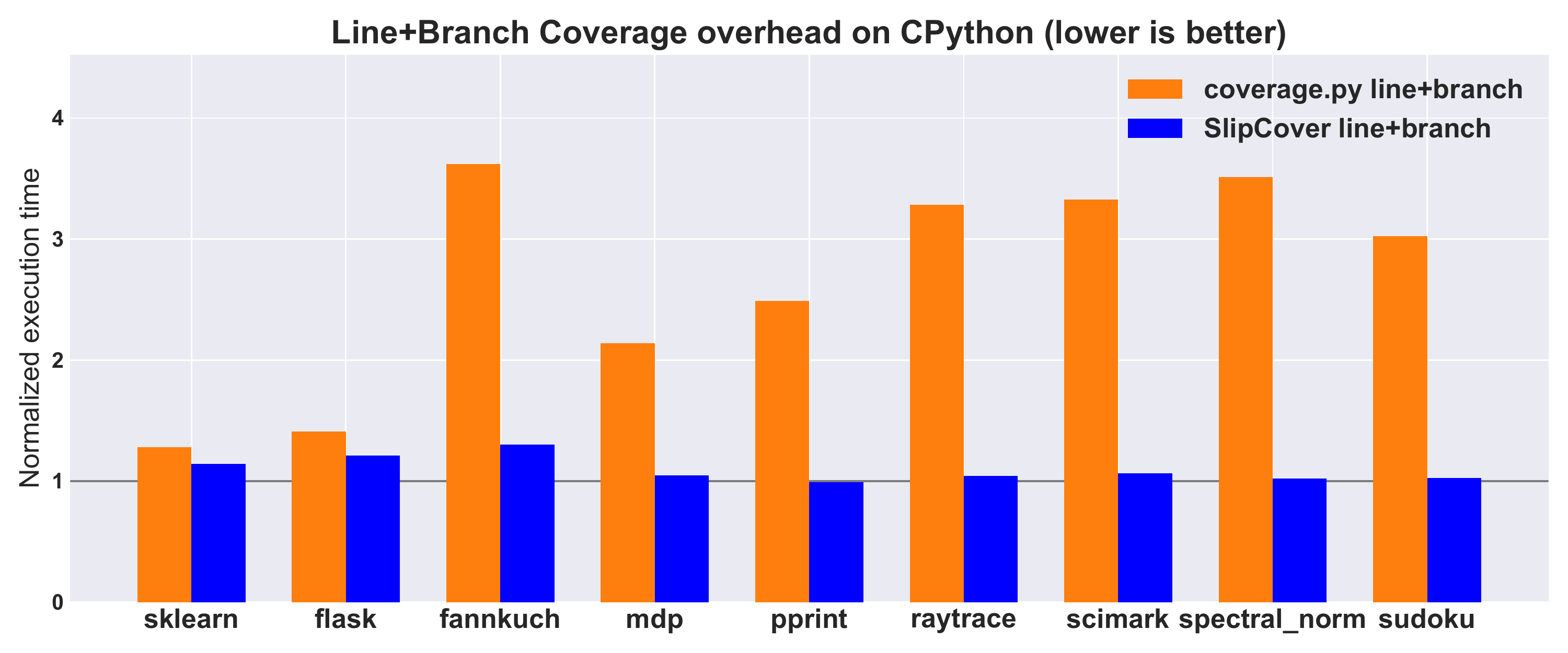}
    \caption{\textbf{[RQ1] \systemname{} is much faster than the state of the art (\texttt{coverage.py}) on CPython:}
    across our benchmark suite, while \texttt{coverage.py} incurs up
    to 260\% overhead, \systemname{}'s median overhead remains at 5\%.
    \label{fig:benchmarks} }
\end{figure}
\begin{figure}[!t]
    \includegraphics[width=\columnwidth]{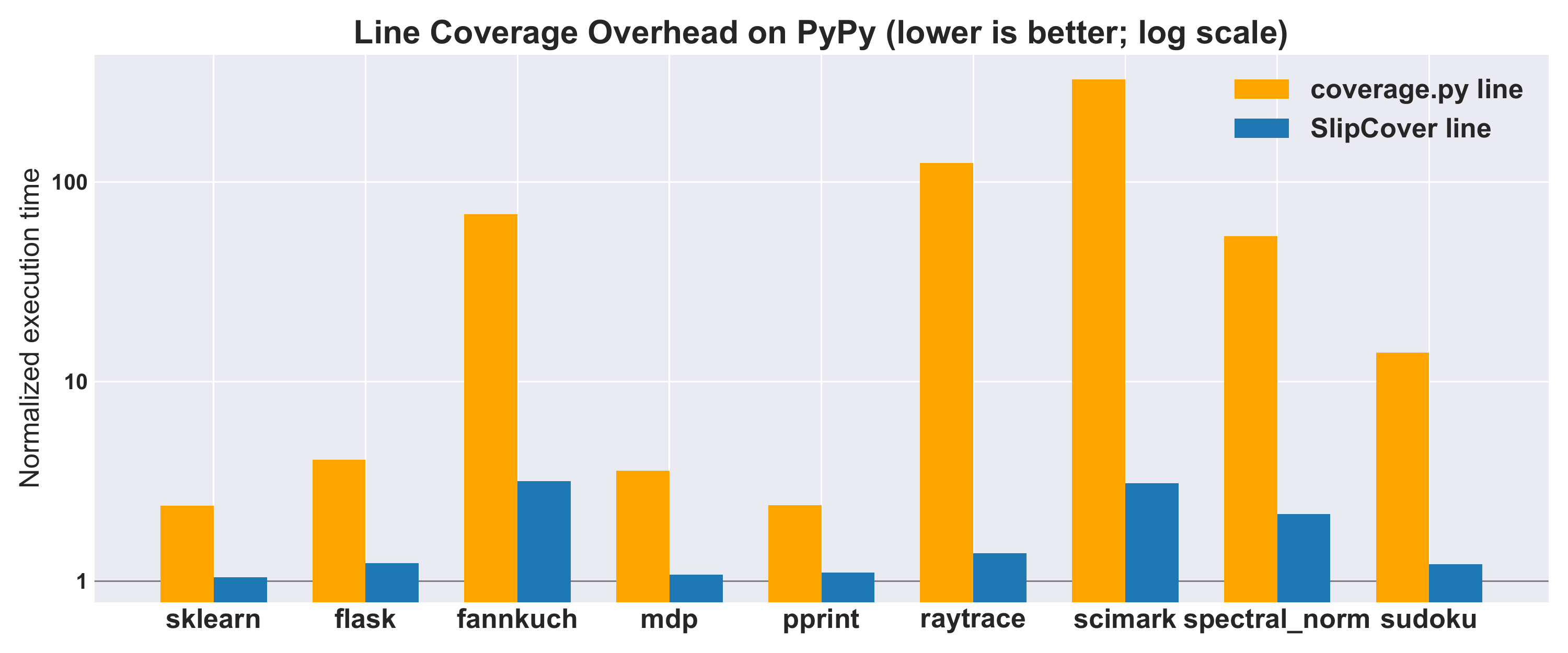}
    \caption{\textbf{[RQ1] \systemname{} is also much faster than \texttt{coverage.py} on PyPy:}
    while \texttt{coverage.py} incurs up to 32,470\% overhead,
    \systemname{}'s median overhead remains at 23\%.
    \label{fig:benchmarks-pypy} }
\end{figure}

Figures~\ref{fig:benchmarks} and~\ref{fig:benchmarks-pypy} present the results of this comparison.
In Figure~\ref{fig:benchmarks}, while \texttt{coverage.py} increases execution times by between
30\% and 260\% (median: 140\%), \systemname{}'s overhead remains at or below 21\% (median: 5\%).
In Figure~\ref{fig:benchmarks-pypy}, while \texttt{coverage.py} increases execution by between
140\% and 32,470\% (median: 140\%), \systemname{}'s overhead remains at or below 220\% (median: 23\%).
PyPy's JIT speeds up the \texttt{raytrace} and \texttt{scimark} benchmarks especially well, leading
to short execution times without coverage analysis.
This leads to \texttt{coverage.py} showing an extreme performance overhead, as the denominator normalizing
that execution is very small.
\textbf{Summary: \systemname{} is much faster than \texttt{coverage.py}
on both CPython and PyPy.}

\subsection{[RQ2] How much does de-instrumentation contribute?}
\label{sec:evaluation-rq2}

As Section~\ref{sec:implementation-de-instrumenting-probes} describes,
\systemname{} de-instruments code in two steps: (1) immediately after
recording new coverage, each probe sets a flag indicating it need not
record it again, and (2) each time a probe is reached, it increments a
counter and eliminates all covered probes when it crosses a threshold.
To quantify the effect of each de-instrumentation step, we
measure \systemname{}'s performance while (a) disabling the probe
elimination, the second step described above, and (b) disabling
de-instrumentation entirely (both steps).

\begin{figure}[!t]
    \includegraphics[width=\columnwidth]{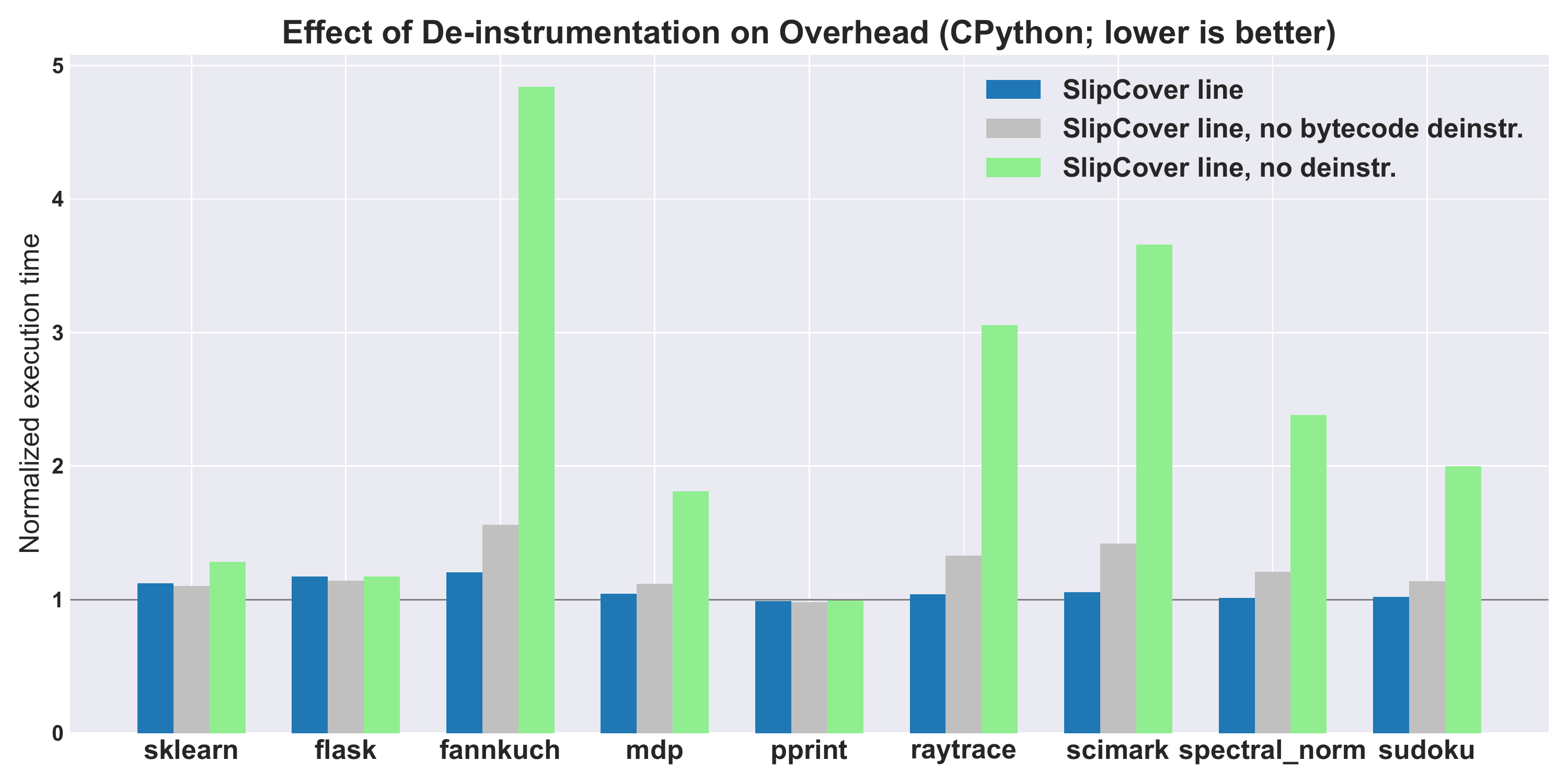} \caption{\textbf{[RQ2]
    De-instrumentation is key to \systemname{}'s performance:} Comparing \systemname{}'s
    normal execution time (blue), to partial de-instrumentation (grey)
    and to no de-instrumentation (green) demonstrates the importance of
    de-instrumentation in reducing \systemname{}'s
    overhead.\label{fig:de-instrumentation} }
\end{figure}

Figure~\ref{fig:de-instrumentation} shows that
simply replacing Python tracing with code instrumentation does not achieve
the performance gains yielded by de-instrumentation.
On the other hand, de-instrumentation is most beneficial to programs that re-execute
previously covered paths, such \texttt{fannkuch} and those on the right of the figure.
The figure also shows that most of the de-instrumentation overhead is eliminated in step (1).
\textbf{Summary: De-instrumentation is key to reducing \systemname{}'s overhead.}

\subsection{[RQ3] How does \systemname{}'s approach behave across Python versions?}
\label{sec:evaluation-rq3}

Python is currently the subject of efforts to increase its
performance~\cite{faster-cpython-3.11, specializing-interpreter}.
It is a natural question to ask how these efforts might
impact the relative performance of \systemname{}
and \texttt{coverage.py}. If the trends show a narrowing gap, then the
long-term benefit of \systemname{} would be unclear.

\begin{figure}[!t]
    \includegraphics[width=\columnwidth]{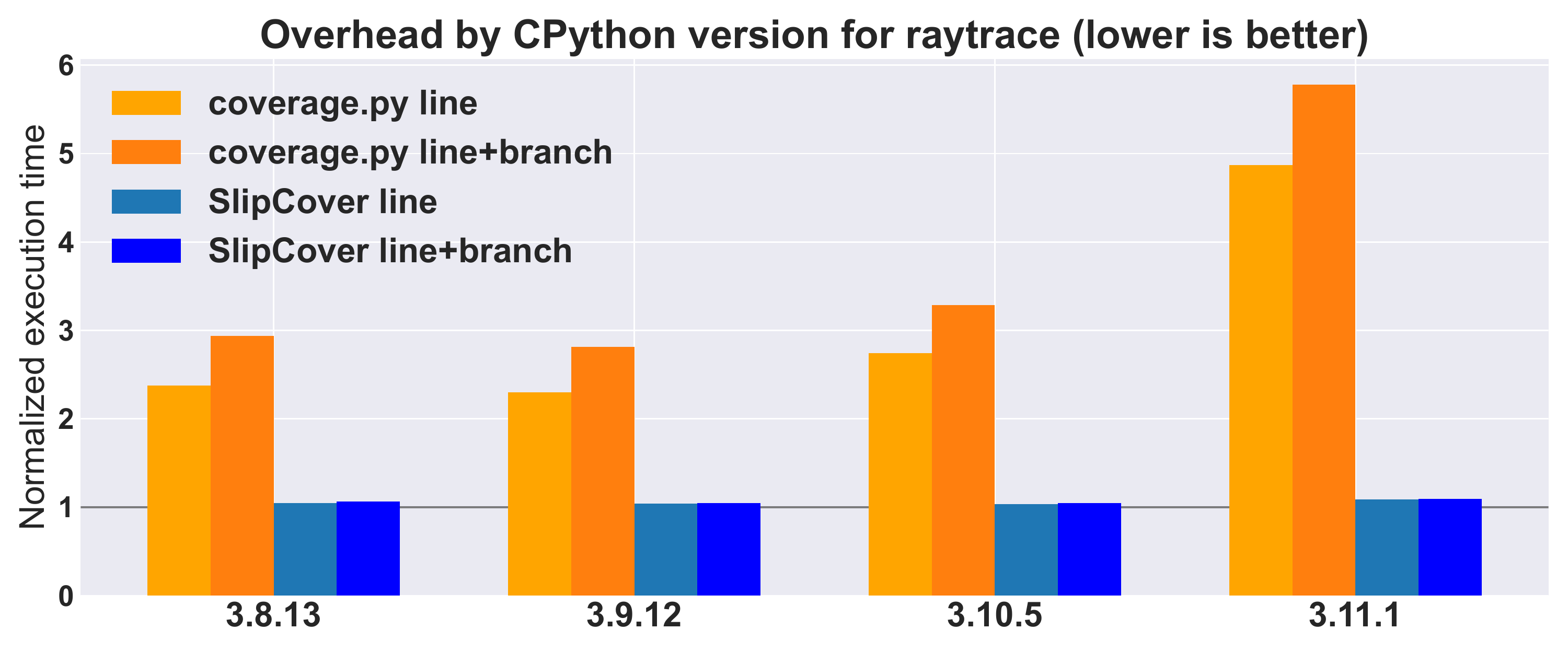}
    \caption{\textbf{[RQ3] \texttt{coverage.py}'s
    relative overhead is growing with successive Python versions,
    while \systemname{}'s overhead remains near zero.} As Python
    developers improve the speed of Python, the added overhead
    of \texttt{coverage.py} consumes an increasingly large portion of
    execution time. \label{fig:overhead-by-version} }
\end{figure}

Figure~\ref{fig:overhead-by-version} compares \texttt{coverage.py}'s
and \systemname{}'s execution times relative to the uninstrumented base case,
without coverage analysis, for four CPython versions, from 3.8 to 3.11.
To simplify the figure, we include only one benchmark, \texttt{raytrace}.

As the graph shows, \texttt{coverage.py}'s performance suffers slightly in
CPython~3.10, and more pronouncedly in CPython~3.11.1 despite developer's
efforts~\cite{python-3_11-tracing-performance-degradation}.
In fact, since execution times are normalized by the time
without coverage analysis, any improvement in Python's
performance reduces the denominator, relatively increasing \texttt{coverage.py}'s
overhead.
\textbf{Summary: \systemname{}'s overhead is near zero across Python versions, while \texttt{coverage.py}'s overhead is growing over time.}

\subsection{[RQ4] What is the overhead of approaches based on Python's tracing mechanism (as used by \texttt{coverage.py})?}
\label{sec:evaluation-rq4}

To investigate whether a new approach to collecting coverage information is
necessary, we create a ``null'' tracer.
Like the one employed by \texttt{coverage.py}, our null tracer registers a
tracing callback using Python's C API~\cite{python-c-tracing-api}.
It also attempts to avoid unnecessary overhead by dynamically turning tracing
off for functions whose coverage information is irrelevant.
These are typically Python library functions, but test suite developers also
commonly specify directories containing relevant code.
The null tracer does not, however, record any lines or branches, or perform
any complicated checks.
Its dynamic tracing control only checks the source file path prefix.

\begin{figure}[!t]
    \includegraphics[width=\columnwidth]{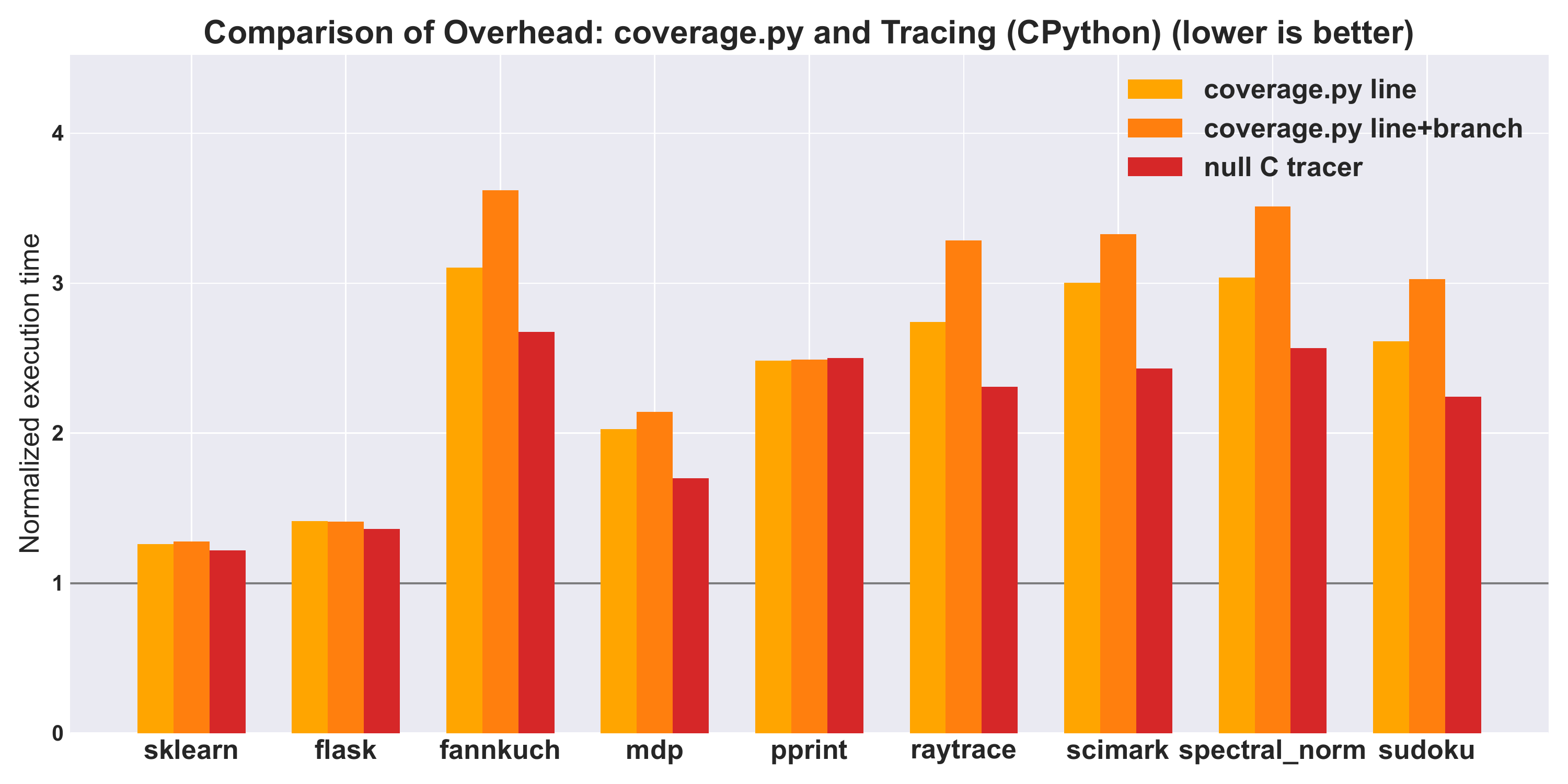}
    \caption{\textbf{[RQ4] Tracing accounts for much of the execution time overhead
    of \texttt{coverage.py}:} the relatively large overhead incurred
    by the ``null'' tracer ($\S$\ref{sec:evaluation-rq4}) strongly suggests that much
    of \texttt{coverage.py}'s running time is, in fact, due to
    tracing.  \label{fig:tracing-overhead} }
\end{figure}

Figure~\ref{fig:tracing-overhead} shows the null tracer's relative running times
next to \texttt{coverage.py}'s.
The null tracer's overhead ranges between 42\% and 99\% (median: 63\%) of
\texttt{coverage.py}'s overhead. \textbf{Summary: tracing is expensive, and comprises much of the cost of \texttt{coverage.py}.}

\subsection{[RQ5]: What is the impact of de-instrumentation when running in the PyPy JIT compiler?}
\label{sec:evaluation-rq5}

\begin{figure}[!t]
    \includegraphics[width=\columnwidth]{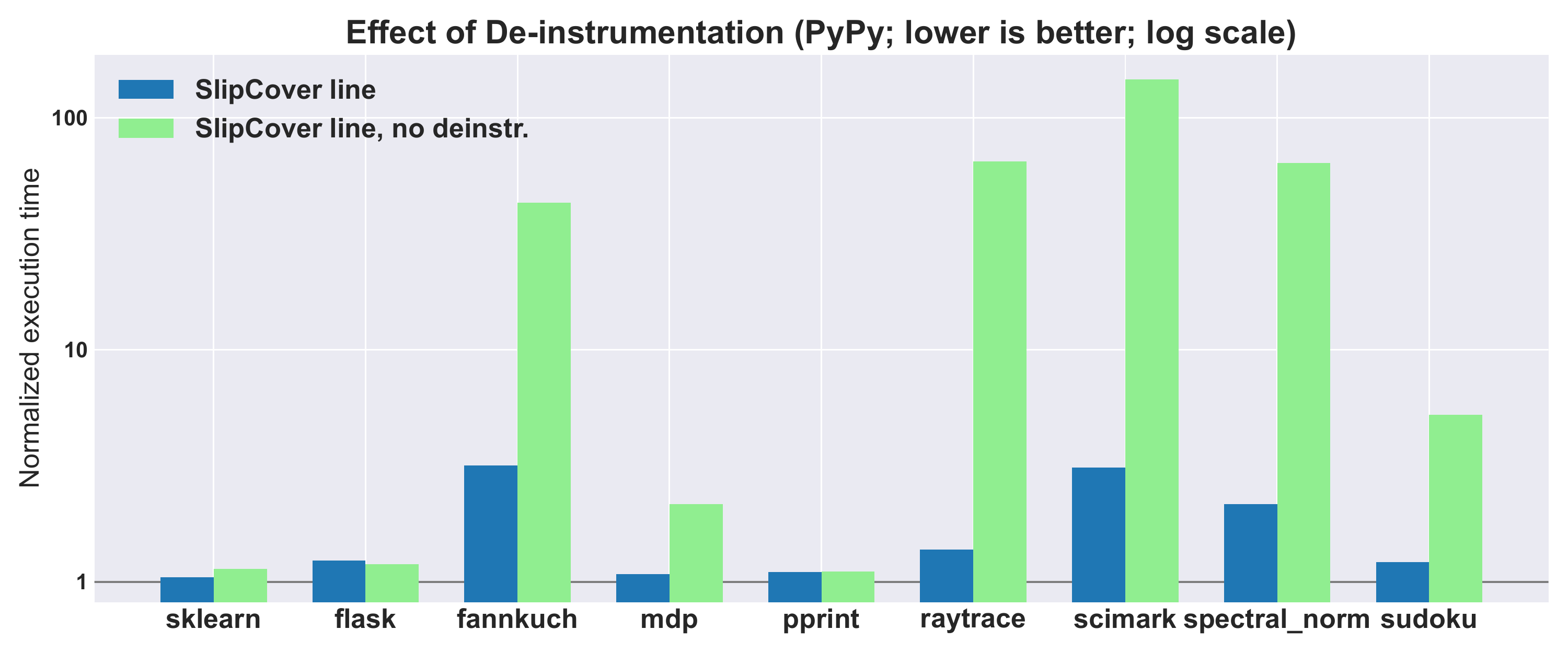}
    \caption{\textbf{[RQ5] De-instrumentation on PyPy}:
        de-instru\-mentation is vital for \systemname{}'s performance for some benchmarks on PyPy,
        despite its JIT compiler.
        \label{fig:pypy-tracing-overhead}
    }
\end{figure}

The effective de-instrumentation we observe through the Java JIT compiler optimizations
(Section~\ref{sec:java-jit}) raises the question whether \systemname{}'s
de-instrumentation is still relevant in the context of PyPy, a JIT compiler for Python.
To answer this question, we compare the overhead of \systemname{}'s coverage analysis
to its overhead if all de-instrumentation is disabled.
Figure~\ref{fig:pypy-tracing-overhead} shows the results. 
While de-instrumentation does not significantly improve performance
for all benchmarks, it dramatically reduces overhead for several. \textbf{Summary: de-instrumentation has a dramatic impact on performance when running with PyPy's JIT compiler, reducing overhead by $40\times$--$150\times$.}

\subsection{[RQ6]: How does \systemname{} affect the performance of automated testing frameworks?}
\label{sec:evaluation-rq6}
As a proof of concept, we utilize an example program from TPBT~\cite{tpbt},
a property-based testing system, that originally draws coverage information
from \texttt{coverage.py} to guide the testing.
We create an alternative version that measures coverage using \systemname{}
and compare their running times.
We observe that with \systemname{}, the program runs in~1s,
about 22$\times$ faster than the 22.3s it runs in when drawing
coverage information from \texttt{coverage.py}.
\textbf{Summary: \systemname{} can greatly improve the performance of
coverage-guided testing.}

  \section{Other Programming Languages}
  \label{sec:portability}
Many of the techniques \systemname{} demonstrates in the context of Python
should be applicable to other bytecode-based languages.
These may naturally require some adaptation depending on what is available
or efficient in the language and execution environment.
While evaluating this in detail goes beyond the scope of this paper,
we explore the question of applicability to other programming languages by creating and evaluating a proof-of-concept
version of \systemname{} for Java.

Prior work for Java~\cite{DBLP:journals/stvr/ChilakamarriE06} required changes to the
JVM because at the time no API was available to allow bytecode modification at runtime;
we instead utilize the Java Instrumentation API~\cite{java-instrumentation}.

\subsubsection*{De-instrumentation cost}
When on-stack bytecode replacement is not supported by the interpreter,
de-instrumenting it incurs a larger cost.
In Python, such cost stems from replacing the bytecode object and
updating references to it (Section~\ref{sec:implementation-de-instrumenting-probes}).
In Java, the Instrumentation API requires the modified bytecode to be passed
in class file format. The cost then stems from recreating the entire class containing the
changed bytecode, as well as loading and eventually JIT-recompiling it.

In such situations, it becomes important to manage these costs, only de-instrumenting
bytecode that seems likely to execute again.
While \systemname{}'s simple per-probe threshold proved effective for
Python, initial data for Java suggests that modifying the mechanism
to a per-method or per-class threshold might better balance de-instrumentation
gains and costs given that the Instrumentation API modifies an entire class at once.


\subsubsection*{Lightweight probes}
A probe might execute multiple times until its de-instrumentation is triggered
and the replacement bytecode is used.
For that reason, it is important to keep probes lightweight, that is,
incurring as low overhead as possible.
\systemname{}'s Python probes were designed as a bytecode and native hybrid
because in Python (and especially in CPython, where all execution is
interpreted), using native code is crucial to performance.
In Java, our prototype utilizes static boolean arrays and method-local variables
created during instrumentation to keep overhead as low as possible.

\subsubsection*{Optimizing JIT compilers}
\label{sec:java-jit}
We compare the prototype's running time with that of a popular Java
coverage analyzer, JaCoCo~\cite{jacoco}. We discover to our surprise
that JaCoCo's performance rivals and in certain cases even exceeds the
prototype's performance.

To investigate why, we instrument a small Java program with JaCoCo.
Figure~\ref{fig:java-program} shows the program, which contains a loop that repeatedly
divides a variable by a constant (line~5) and adds a constant to it (line~6).

\begin{figure}[!t]
\begin{lstlisting}[language=Java,numbers=left,xleftmargin=10pt]
    class T {
        static void foo() {
            double n = 0.2;
            for (int i=0; i<100000; i++) {
                n /= 1.1;
                n += .1;
            }
            System.out.println("n=" + n);
        }
    
        public static void main(String... args) {
            foo();
        }
    }
\end{lstlisting}
    \caption{\textbf{Sample Java program:}
        we instrument this program with JaCoCo and review the native code
        emitted by the JVM's JIT compiler to understand how JaCoCo achieves high performance despite not performing de-instrumentation.
        \label{fig:java-program}
    }
\end{figure}

Figure~\ref{fig:java-bytecode} shows a portion of the resulting Java bytecode, 
highlighting the JaCoCo instrumentation.
JaCoCo records coverage by creating a Boolean array where each element corresponds to
a portion of the code.
Bytecode offsets 0--5 in Figure~\ref{fig:java-bytecode} provide access to that array.
JaCoCo also inserts various probes into bytecode that set elements in the array to \texttt{true}
as they execute.
Bytecode offsets 12--15 contain such a probe.
\begin{figure}[!t]
\begin{lstlisting}
 0: $\textbf{ldc}$        $\textbf{\#65}$  // Dynamic #1:$\$$jacocoData:Ljava/lang/Object;
 2: $\textbf{checkcast}$   $\textbf{\#67}$  // class "[Z"
 5: $\textbf{astore\_0}$
 6: ldc2_w     #7
 9: dstore_1
10: iconst_0
11: istore_3
12: $\textbf{aload\_0}$
13: $\textbf{iconst\_1}$
14: $\textbf{iconst\_1}$
15: $\textbf{bastore}$
16: iload_3
17: ldc        #9
19: if_icmpge  44
[...]
\end{lstlisting}
    \caption{\textbf{JaCoCo instrumented bytecode:} 
        This bytecode results from compiling the sample program in Figure~\ref{fig:java-program}
        and instrumenting it with JaCoCo.
        At the beginning of the method, JaCoCo creates a Boolean array to record coverage.
        It also inserts probes that set array elements to \texttt{true}.
        \label{fig:java-bytecode}
    }
\end{figure}

After instrumenting the program, we execute it by invoking the
JVM using the following flags:\texttt{-XX:+UnlockDiagnosticVMOptions -XX:CompileCommand=print,T.foo}.
which logs native code produced by its JIT compiler, if any.

Figure~\ref{fig:java-disassembly} shows a portion of the x64 code logged by the JVM.
At the top of the listing, a \texttt{movb} instruction stores the value \texttt{1}.
That corresponds to assigning \texttt{true} to an element of JaCoCo's coverage Boolean array.
The \texttt{vdivsd} and \texttt{vaddsd} instructions implement the division and addition
operations from source code lines 5 and 6.

The JIT compiler thus optimized the loop, unrolling it, and also hoisted any Boolean
assignments out of it. The loop consists exclusively of the arithmetic instructions
already mentioned, followed by an iteration counter update (adding \texttt{0x10} at a
time, as the loop was unrolled to perform 16 operations without further checking).
\begin{figure}[!t]
\begin{lstlisting}
=================== C2-compiled method ====================
[...]
0x00007f198d1825f5:   movb   $\$$0x1,0x12(%r13)
[...]
0x00007f198d182630:   vdivsd -0x118(%rip),%xmm0,%xmm0
0x00007f198d182638:   vaddsd -0x118(%rip),%xmm0,%xmm0
0x00007f198d182640:   vdivsd -0x128(%rip),%xmm0,%xmm0
0x00007f198d182648:   vaddsd -0x128(%rip),%xmm0,%xmm0
0x00007f198d182650:   vdivsd -0x138(%rip),%xmm0,%xmm0
0x00007f198d182658:   vaddsd -0x138(%rip),%xmm0,%xmm0
[...]
0x00007f198d182720:   vdivsd -0x208(%rip),%xmm0,%xmm0
0x00007f198d182728:   vaddsd -0x208(%rip),%xmm0,%xmm0
0x00007f198d182730:   add    $\$$0x10,%ebp
0x00007f198d182733:   cmp    %r10d,%ebp
0x00007f198d182736:   jl     0x00007f198d182630
[...]
\end{lstlisting}
    \caption{\textbf{JIT generated code omits instrumentation:}
        Java's C2 JIT compiler generated this x64 code, unrolling
        the sample program's loop and omitting (hoisting) the \texttt{movb}
        instructions used to record coverage from it.
        \label{fig:java-disassembly}
    }
\end{figure}

The result is that in modern implementations of Java, the optimizing
JIT compiler eliminates much of the overhead of probes, even without
de-instrumentation.
The limited remaining performance gap does not justify the effort of producing a full port of \systemname{} to Java.

Section~\ref{sec:evaluation-rq5} shows that the
same situation does not apply to Python. When running with the PyPy
JIT, not de-instrumenting can in some cases degrade
performance dramatically, slowing execution by $40\times$--$150\times$.

  \section{Prior Work}
  \label{sec:related-work}
This section describes previous work aimed at reducing the overhead of
coverage analysis.  Section~\ref{sec:static-instrumentation} first
describes approaches that instrument code only statically (entirely
prior to execution time). Section~\ref{sec:dynamic-instrumentation}
describes approaches that (also) modify code dynamically (during
execution).
Section~\ref{sec:comparison} compares \systemname{}'s design,
algorithms and implementations with past approaches.
Finally, Section~\ref{sec:related-coverage-comparison}
details \texttt{coverage.py}, the current state-of-the-art profiler
for Python.

\subsection{Static instrumentation}
\label{sec:static-instrumentation}

\subsubsection*{Java}

As far as we are aware, \citet{residual} are the first to propose
removing instrumentation once coverage is recorded. Their approach,
which the paper calls ``residual'' test coverage, is to instrument
bytecode in Java class files to record when each basic block is
reached, skipping those that a previously-instrumented execution had
reached (each execution updates the class files on disk). It works
with stock (unmodified) JVMs but is only able to de-instrument code
the next time the program executes.

\subsubsection*{Native code}

\citet{DBLP:conf/sp/NagyH19} present a similar approach for use in fuzzing called ``full speed fuzzing.''
It performs binary instrumentation to trigger software interrupts whenever basic blocks are reached for the first time.
It rewrites the binary on disk to de-instrument that basic block,
and then re-executes the binary to continue fuzzing.

\subsection{Dynamic instrumentation}
\label{sec:dynamic-instrumentation}

\subsubsection*{Native code}

\citet{DBLP:journals/jss/TikirH05} implement a code coverage analyzer by
extending DyninstAPI~\cite{DBLP:journals/ijhpca/BuckH00}, a library for dynamic
native code instrumentation.
It is the earliest work to both dynamically insert and remove coverage probes
during program execution. It is limited to line coverage.
To reduce the initial overhead of instrumentation, it performs control-flow dominator
based static analysis on the code and only places probes in certain basic blocks.
It implements both pre-instrumentation, which inserts all probes ahead of time,
and on-demand instrumentation, which inserts breakpoints at the beginning of
functions to perform the required analysis and then insert probes when a
function is executed. Finally, it periodically (at fixed time intervals) removes
probes that are no longer needed.

\subsubsection*{Java}

\citet{DBLP:journals/stvr/ChilakamarriE06} proposes a similar ``disposable'' coverage instrumentation for Java.
It instruments JVM bytecode and, once a probe is no longer needed,
de-instrument it by overwriting the probe with \texttt{NOP}
operations.  It requires modifications to the JVM to trigger
de-instrumentation of both interpreted and JIT-compiled bytecode. We
note that it might be possible today to accomplish this via the Java
Instrumentation API~\cite{java-instrumentation}, which was not
available at that time. Section~\ref{sec:java-jit} presents an evaluation
of a prototype using this approach.

Another Java-based approach, \citet{Misurda1553558} implements both
dynamic probe insertion and removal. In the same vein as
\citet{DBLP:journals/jss/TikirH05}, it only pre-instruments by
inserting \emph{seed} probes which further instrument basic blocks
once reached. Its approach is to instrument the x86 code generated by
the JIT compiler, relying on support provided by Jikes, a research
virtual machine. This approach would be unsound in modern,
industrial-strength JVM implementations, which initially interpret
bytecode.

Surprisingly, as Section~\ref{sec:java-jit} shows, the
optimizations that modern Java JIT compilers perform eliminate most of
the overhead of instrumented code, yielding few opportunities to gain
performance by de-instrumentation in the context of Java.


\subsection{Comparisons with \systemname{}}
\label{sec:comparison}

This section characterizes the differences with prior work by
contrasting \systemname{}'s algorithmic and technical approaches.

\subsubsection*{On-the-fly de-instrumentation}

While Python does not require compilation into bytecode ahead of time,
it caches previous compilations in \texttt{.pyc} files as a way to
speed up program startup.  While it would be possible to adapt the
above-described approaches to Python bytecodes, it would still delay
all de-instrumentation until a subsequent execution (if any).  By
contrast, \systemname{} de-instruments code live, immediately speeding
the current program execution.

\subsubsection*{Aggressive vs. optimized probe insertion}

\systemname{} does not attempt to reduce the number of probes
inserted, but instead adds probes for every line and branch ahead of
program execution. \systemname{} adopts this design decision based on the
observation of \cite{DBLP:journals/jss/TikirH05}: ``the most
significant gains come from removing instrumentation code [...] rather
than from binary analysis algorithms to optimize instrumentation
placement.'' In fact, such an approach that adds bytecode probes on
demand would be difficult to adapt to Python, as bytecode objects are
read-only.

Beyond that technical issue, the approach of attempting to avoid
overhead \emph{a priori} is complicated by the fact that in Python,
many opcodes can result in a branch by throwing an exception. The
result would be basic blocks consisting of few or even just a single opcode.

\subsubsection*{Stepwise de-instrumentation}

As Section~\ref{sec:implementation-de-instrumenting-probes} describes, \systemname{} reduces probe
overhead in two steps: one immediately once the probe is no longer
needed, and another threshold-based, thus amortizing the cost of
de-instrumentation.  By
contrast, \citet{DBLP:journals/jss/TikirH05}'s approach only removes probes at
fixed time intervals, which may be either too aggressive or too lazy.

Overwriting a probe with \texttt{NOP}
operations~\cite{DBLP:journals/stvr/ChilakamarriE06} is convenient for
Java because it does not require updating any bytecode metadata.  If a
Java application were to add a jump opcode, it would need to update
the bytecode's \texttt{StackMapTable}~\cite{stackmaptable}, a more
complex operation.  Python does not have the same restriction,
so \systemname{}'s second de-instrumentation step uses a jump opcode
to effectively remove probes.

\subsubsection*{No modifications to VMs}

\systemname{} executes on unmodified Python interpreters.
Approaches like \citet{DBLP:journals/stvr/ChilakamarriE06} that require
modifying the VM are undesirable in many settings: a modified Python
interpreter or JIT would require separate installation and likely
hamper adoption.

\subsubsection*{Portable and complete instrumentation}

While operating at the x86 level lets \citet{Misurda1553558} benefit
from efficient code manipulation (while sacrificing portability
to non-x86 platforms), it also makes its approach generally unsuitable
for Python. Despite exceptions such as native extension modules and
compiled Python variants, Python typically is interpreted (entirely in
CPython, and initially in PyPy). \citet{Misurda1553558}'s approach
would also be unsound in any modern Java implementation, which only
JIT-compile code once the interpreter has executed the bytecode long
enough to identify its hot spots. By contrast, \systemname{} ensures
complete coverage of all Python byte code.



\subsection{Comparison with \texttt{coverage.py}}
\label{sec:related-coverage-comparison}

\texttt{coverage.py}~\cite{coveragepy} is the standard
coverage analysis tool for Python.  It uses Python's profiling and
tracing API~\cite{python-c-tracing-api} to collect dynamic coverage
information, registering \textit{tracer} functions that Python invokes
whenever it reaches a line.  Section~\ref{sec:evaluation-rq4} shows that using
this API introduces significant overhead, accounting for much of
the \texttt{coverage.py}'s high overhead. \systemname{}'s use of
bytecode instrumentation and de-instrumentation avoids this overhead.

\texttt{coverage.py} collects strictly less information than \systemname{}.
Rather than performing line coverage, \texttt{coverage.py}
performs \emph{statement} coverage. Python statements can easily span
multiple lines, including whenever there is an unclosed bracket or
parenthesis, or by placing backslashes at the end of each
line. \texttt{coverage.py} performs a pre-processing stage that omits
continued lines when computing the set of all lines in the source.

For example, \texttt{coverage.py} does not report line~5 as not
covered in the code below. 
By contrast,
\systemname{} correctly indicates that line~5 is not covered.

\begin{lstlisting}[language=Python,numbers=left,xleftmargin=10pt]
    cond = True
    foo = {
        "const": 42,
        "computed": 666 if cond \
                    else 999,
        "other": 0
    }
\end{lstlisting}

An analogous problem arises with branch coverage. \texttt{coverage.py}
detects branches by recording \textit{arcs}, which are line
transitions it observes as the program execution moves from one line
to another.  Unfortunately, this method prevents
\texttt{coverage.py} from detecting same-line branches. Consider the following code:

\begin{lstlisting}[language=Python]
    if x > 0: print("positive")
\end{lstlisting}

Because \texttt{coverage.py} sees no line transitions, it incorrectly
does not detect the branch in this code.
Unlike \texttt{coverage.py}, \systemname{}'s AST-based instrumentation
allows it to correctly detect all branches.


  \section{Conclusion}
  \label{sec:conclusion}

This paper presents \systemname{}, a novel coverage analyzer
that employs a combination of AST transformations and dynamic bytecode instrumentation
and de-instrumentation to reduce the cost of collecting branch and line coverage information.
\systemname{} works with both the Python interpreter and PyPy, without modifications.
It shows that bytecode de-instrumentation can be effective in industrial-strength language
implementations, even when on-stack bytecode replacement is not supported.

Compared to \texttt{coverage.py},
the state-of-the-art Python coverage analyzer, \systemname{} reduces overhead from 30\%--260\% to 5\% when
running with the Python interpreter, and from 140\%--32,400\% to 23\% when running with PyPy.
Using \systemname{} can speed up coverage-guided testing by 22$\times$.
\systemname{} has been released as open source at
\ifanonymous{(URL removed for double-blind review)}%
\ifnotanonymous{\url{https://github.com/plasma-umass/slipcover}}.

  \makeatletter
  \if@ACM@anonymous

  \else
      \section*{Acknowledgements}
      We would like to thank Ned Batchelder for answering our questions about \texttt{coverage.py}
      and for his  support for this work.
      This upon work has been supported by the National Science Foundation under Grant No.~1955610.
      Any opinions, findings, and conclusions or recommendations expressed in this material are those
      of the author(s) and do not necessarily reflect the views of the National Science Foundation.

  \fi
  \makeatother

  \appendix
  \section{Evaluation Data}
  This appendix shows the evaluation data from Section~\ref{sec:evaluation}.

\setlength{\abovecaptionskip}{3pt}

\renewcommand{\theadfont}{\fontsize{6}{7}\selectfont}

\begin{table}[H]
    \caption{[RQ1] Data for Figure~\ref{fig:benchmarks},
        showing each benchmark's execution time without coverage analysis,
        with line and line+branch coverage analysis by both
        \texttt{coverage.py} and \systemname{} running on CPython.
    }
    \small
    \begin{tabularx}{\columnwidth}{@{} X r r r r r @{}}
\thead[bl]{Benchmark} & \thead[br]{no coverage} & \thead[br]{\texttt{coverage.py}\\line} & \thead[br]{\texttt{coverage.py}\\line+branch} & \thead[br]{\systemname{}\\line} & \thead[br]{\systemname{}\\line+branch} \\
\hline
\texttt{sklearn} & \SI{527.8}{s} & \SI{666.1}{s} & \SI{675.0}{s} & \SI{592.8}{s} & \SI{603.3}{s} \\
\texttt{flask} & \SI{8.0}{s} & \SI{11.3}{s} & \SI{11.3}{s} & \SI{9.4}{s} & \SI{9.7}{s} \\
\texttt{fannkuch} & \SI{15.1}{s} & \SI{46.9}{s} & \SI{54.6}{s} & \SI{18.2}{s} & \SI{19.7}{s} \\
\texttt{mdp} & \SI{10.4}{s} & \SI{21.1}{s} & \SI{22.3}{s} & \SI{10.9}{s} & \SI{10.9}{s} \\
\texttt{pprint} & \SI{13.0}{s} & \SI{32.3}{s} & \SI{32.4}{s} & \SI{12.9}{s} & \SI{12.9}{s} \\
\texttt{raytrace} & \SI{9.5}{s} & \SI{26.1}{s} & \SI{31.2}{s} & \SI{9.9}{s} & \SI{9.9}{s} \\
\texttt{scimark} & \SI{8.2}{s} & \SI{24.5}{s} & \SI{27.1}{s} & \SI{8.6}{s} & \SI{8.7}{s} \\
\texttt{spectral\_norm} & \SI{8.1}{s} & \SI{24.7}{s} & \SI{28.6}{s} & \SI{8.3}{s} & \SI{8.3}{s} \\
\texttt{sudoku} & \SI{7.9}{s} & \SI{20.8}{s} & \SI{24.1}{s} & \SI{8.1}{s} & \SI{8.2}{s} \\
\hline
\end{tabularx}
\end{table}

\begin{table}[H]
    \caption{[RQ1] Data for Figure~\ref{fig:benchmarks-pypy}.
        showing each benchmark's execution time without coverage analysis
        and with line coverage analysis by both
        \texttt{coverage.py} and \systemname{} running on PyPy.
    }
    \small
\begin{tabularx}{\columnwidth}{@{} X r r r @{}}
\thead[bl]{Benchmark} & \thead[br]{no coverage} & \thead[br]{\texttt{coverage.py} line} & \thead[br]{\systemname{} line} \\
\hline
\texttt{sklearn} & \SI{2910.2}{s} & \SI{6970.0}{s} & \SI{3045.9}{s} \\
\texttt{flask} & \SI{15.9}{s} & \SI{64.8}{s} & \SI{19.7}{s} \\
\texttt{fannkuch} & \SI{3.2}{s} & \SI{223.9}{s} & \SI{10.3}{s} \\
\texttt{mdp} & \SI{17.4}{s} & \SI{62.3}{s} & \SI{18.7}{s} \\
\texttt{pprint} & \SI{2.6}{s} & \SI{6.3}{s} & \SI{2.9}{s} \\
\texttt{raytrace} & \SI{0.7}{s} & \SI{88.3}{s} & \SI{1.0}{s} \\
\texttt{scimark} & \SI{0.3}{s} & \SI{108.6}{s} & \SI{1.0}{s} \\
\texttt{spectral\_norm} & \SI{0.4}{s} & \SI{22.6}{s} & \SI{0.9}{s} \\
\texttt{sudoku} & \SI{4.6}{s} & \SI{64.1}{s} & \SI{5.6}{s} \\
\hline
\end{tabularx}
\end{table}

\begin{table}[H]
    \caption{[RQ2] Data for Figure~\ref{fig:de-instrumentation},
        showing the effect of disabling \systemname{}'s de-instrumentation
        steps on line coverage analysis, running on CPython.
    }
    \small
\begin{tabularx}{\columnwidth}{@{} X r r r r @{}}
\thead[bl]{Benchmark} & \thead[br]{no coverage} & \thead[br]{\systemname{} line} & \thead[br]{\systemname{} line,\\  no bytecode\\de-instr.} & \thead[br]{\systemname{} line,\\  no de-instr.} \\
\hline
\texttt{sklearn} & \SI{527.8}{s} & \SI{592.8}{s} & \SI{581.5}{s} & \SI{677.2}{s} \\
\texttt{flask} & \SI{8.0}{s} & \SI{9.4}{s} & \SI{9.2}{s} & \SI{9.4}{s} \\
\texttt{fannkuch} & \SI{15.1}{s} & \SI{18.2}{s} & \SI{23.6}{s} & \SI{73.1}{s} \\
\texttt{mdp} & \SI{10.4}{s} & \SI{10.9}{s} & \SI{11.6}{s} & \SI{18.9}{s} \\
\texttt{pprint} & \SI{13.0}{s} & \SI{12.9}{s} & \SI{12.8}{s} & \SI{12.9}{s} \\
\texttt{raytrace} & \SI{9.5}{s} & \SI{9.9}{s} & \SI{12.6}{s} & \SI{29.1}{s} \\
\texttt{scimark} & \SI{8.2}{s} & \SI{8.6}{s} & \SI{11.6}{s} & \SI{29.8}{s} \\
\texttt{spectral\_norm} & \SI{8.1}{s} & \SI{8.3}{s} & \SI{9.8}{s} & \SI{19.4}{s} \\
\texttt{sudoku} & \SI{7.9}{s} & \SI{8.1}{s} & \SI{9.0}{s} & \SI{15.9}{s} \\
\hline
\end{tabularx}
\end{table}

\begin{table}[H]
    \caption{[RQ3] Data for Figure~\ref{fig:overhead-by-version},
        showing \texttt{coverage.py} and \systemname{}'s coverage
        overhead changes with recent CPython versions.
    }
    \small
\begin{tabularx}{\columnwidth}{@{} X r r r r r @{}}
\thead[bl]{Python\\version} & \thead[br]{no coverage} & \thead[br]{\texttt{coverage.py} line} & \thead[br]{\texttt{coverage.py}\\ line+branch} & \thead[br]{\\\systemname{} line} & \thead[br]{\systemname{}\\line+branch} \\
\hline
\texttt{3.8.13} & \SI{9.4}{s} & \SI{22.5}{s} & \SI{27.8}{s} & \SI{9.9}{s} & \SI{10.1}{s} \\
\texttt{3.9.12} & \SI{9.8}{s} & \SI{22.6}{s} & \SI{27.7}{s} & \SI{10.3}{s} & \SI{10.3}{s} \\
\texttt{3.10.5} & \SI{9.5}{s} & \SI{26.1}{s} & \SI{31.2}{s} & \SI{9.9}{s} & \SI{9.9}{s} \\
\texttt{3.11.1} & \SI{5.5}{s} & \SI{26.7}{s} & \SI{31.7}{s} & \SI{6.0}{s} & \SI{6.0}{s} \\
\hline
\end{tabularx}
\end{table}

\begin{table}[H]
    \caption{[RQ4] Data for Figure~\ref{fig:tracing-overhead},
        showing how a ``null'' tracer that activates Python tracing
        compares to \texttt{coverage.py}'s line and line+branch
        coverage overhead on CPython.
    }
    \small
\begin{tabularx}{\columnwidth}{@{} X r r r r @{}}
\thead[bl]{Benchmark} & \thead[br]{no coverage} & \thead[br]{\texttt{coverage.py}\\line} & \thead[br]{\texttt{coverage.py}\\line+branch} & \thead[br]{null C tracer} \\
\hline
\texttt{sklearn} & \SI{527.8}{s} & \SI{666.1}{s} & \SI{675.0}{s} & \SI{642.6}{s} \\
\texttt{flask} & \SI{8.0}{s} & \SI{11.3}{s} & \SI{11.3}{s} & \SI{10.9}{s} \\
\texttt{fannkuch} & \SI{15.1}{s} & \SI{46.9}{s} & \SI{54.6}{s} & \SI{40.4}{s} \\
\texttt{mdp} & \SI{10.4}{s} & \SI{21.1}{s} & \SI{22.3}{s} & \SI{17.7}{s} \\
\texttt{pprint} & \SI{13.0}{s} & \SI{32.3}{s} & \SI{32.4}{s} & \SI{32.5}{s} \\
\texttt{raytrace} & \SI{9.5}{s} & \SI{26.1}{s} & \SI{31.2}{s} & \SI{22.0}{s} \\
\texttt{scimark} & \SI{8.2}{s} & \SI{24.5}{s} & \SI{27.1}{s} & \SI{19.8}{s} \\
\texttt{spectral\_norm} & \SI{8.1}{s} & \SI{24.7}{s} & \SI{28.6}{s} & \SI{20.9}{s} \\
\texttt{sudoku} & \SI{7.9}{s} & \SI{20.8}{s} & \SI{24.1}{s} & \SI{17.8}{s} \\
\hline
\end{tabularx}
\end{table}

\begin{table}[H]
    \caption{[RQ5] Data for Figure~\ref{fig:pypy-tracing-overhead},
        showing the effect of disabling \systemname{}'s de-instrumentation
        on PyPy.
    }
    \small
\begin{tabularx}{\columnwidth}{@{} X r r r @{}}
\thead[bl]{Benchmark} & \thead[br]{no coverage} & \thead[br]{\systemname{} line} & \thead[br]{\systemname{} line,\\ no de-instr.} \\
\hline
\texttt{sklearn} & \SI{2910.2}{s} & \SI{3045.9}{s} & \SI{3302.8}{s} \\
\texttt{flask} & \SI{15.9}{s} & \SI{19.7}{s} & \SI{19.0}{s} \\
\texttt{fannkuch} & \SI{3.2}{s} & \SI{10.3}{s} & \SI{139.9}{s} \\
\texttt{mdp} & \SI{17.4}{s} & \SI{18.7}{s} & \SI{37.6}{s} \\
\texttt{pprint} & \SI{2.6}{s} & \SI{2.9}{s} & \SI{2.9}{s} \\
\texttt{raytrace} & \SI{0.7}{s} & \SI{1.0}{s} & \SI{46.1}{s} \\
\texttt{scimark} & \SI{0.3}{s} & \SI{1.0}{s} & \SI{48.8}{s} \\
\texttt{spectral\_norm} & \SI{0.4}{s} & \SI{0.9}{s} & \SI{27.0}{s} \\
\texttt{sudoku} & \SI{4.6}{s} & \SI{5.6}{s} & \SI{24.1}{s} \\
\hline
\end{tabularx}
\end{table}

%
%
%

\punt{
\begin{table}[H]
    \caption{[RQ6] Data for proof of concept comparison between guiding TPBT~\cite{tpbt}
        with \texttt{coverage.py} and \systemname{}.
    }
    \centering
    \small
    \begin{tabular}{l r}
        \thead[l]{Benchmark} & \thead[r]{execution time} \\
        \hline
        \texttt{coverage.py} & 22.3s \\
        \systemname{} & 1.0s \\
        \hline
    \end{tabular}
\end{table}
}

  {
  \balance
  \bibliographystyle{ACM-Reference-Format}
  \bibliography{emery,slipcover}
  }

\end{document}